\documentclass[11pt,twoside]{article}
 \usepackage{./asp2014}

\aspSuppressVolSlug
\resetcounters

\begin{document}

\title{North Galactic plane structure with IPHAS Be stars.}
\author{Gkouvelis L.,  Fabregat J. $\&$ the IPHAS consortium. }
\affil{Observatori Astronom\'ic,Universitat de Valencia, Valencia, Spain; \email{leonardo.gkouvelis@uv.es, juan.fabregat@uv.es}}

\paperauthor{Gkouvelis L.}{leonardo.gkouvelis@uv.es}{ORCID_Or_Blank}{University of Valencia}{Physics Department}{Valencia}{Valencia}{49006}{Spain}

\begin{abstract}
Our goal is to investigate the spiral structure of the Northern Galactic plane using as tracers the classical Be stars detected by INT Photometric H$\alpha$ Survey (IPHAS). IPHAS scans the $29^o<l<+215^o, -5^o<b<+5^o$ region in the $r$,  $i$ and H$\alpha$ bands. Spectroscopic follow up has been done for the bright  H$\alpha$ emitters. We have developed an automatic procedure for spectral analysis, based on the BCD spectrophotometric system. In this paper we present a cataloque of 1135 Classical Be stars, for which we have determined  spectral types, astrophysical parameters and  distances. From these results we make a first attempt to map the structure of the Galactic disk in the anticenter direction.
\end{abstract}

\section{ Introduction}

The INT Photometric H$\alpha$ Survey of the Northern Galactic plane provides imaging photometry in the $r$, $i$ and H$\alpha$ filters for stars up to twentieth magnitude in the region $29^o<l<+215^o, -5^o<b<+5^o$ \citep{Drew2005}. The aim of the survey is to detect objects with emission in the H$\alpha$ line.  The IPHAS colour-colour diagram provides an efficient tool to detect bright H$\alpha$ emitters \citep{Corradi2008}.\par

In this work we present a study of a sample of Classical Be (CBe) stars, which constitute a significant percentage of of the IPHAS detected H$\alpha$ emitters.  We use the CBe stars as tracers of the galactic structure, in order to study the spiral arms. Our aim is to contribute to the scientific debate on the compact structure of the Perseus Arm, the existence of the Outer Arm and how long the latter extends. We present  preliminary results on the distances to 1135 CBe stars, and we discuss on the results.\par 

\section{Analysis}
Spectroscopic follow-up of bright H$\alpha$ emitters detected by IPHAS was carried out at the Fred Lawrence Whipple Observatory with the 1.5m telescope and the FAST spectrograph. Observations started at 2005 and finished at 2012, collecting 2627 spectra. The limits in the $r$ band are between 13 and 17 mag. Almost 70\% of the obtained spectra correspond to CBe stars. We selected among them those with $SNR\geq30$ in the Balmer discontinuity region, to ensure that accurate spectral parameters can be derived. This lead to a final sample of 1169 spectra to study, corresponding to 1135 stars.  For thirty-two  of them we have two spectra, and three spectra for one more star.\par

To analyse this large amount of data we have developed an automatic spectral classification procedure based on the Barbier-Chalonge-Divan (BCD) spectrophotometric system (\cite{Barbier1941}, \cite{Chalonge1952}). The analysis procedure is described in detail in  Gkouvelis et al.  2014 (in preparation). From the spectra we extract the BCD parameters and the H$\alpha$ EW, so that we can determine the spectral classification, astrophysical parameters, absolute magnitude and finally, with the combination of the IPHAS photometry, the distance  for each star.\par

\section{Results} 
We calculated the distances and the astrophysical parameters of 1135 CBe stars in the Northern Galactic plane. In Fig.1 we show the distance of each one, overplotted to a scaled image from Hurt (2008), which illustrates the scientific conclusions of the Spitzer space telescope observations in the Galactic plane.\par

\articlefigure{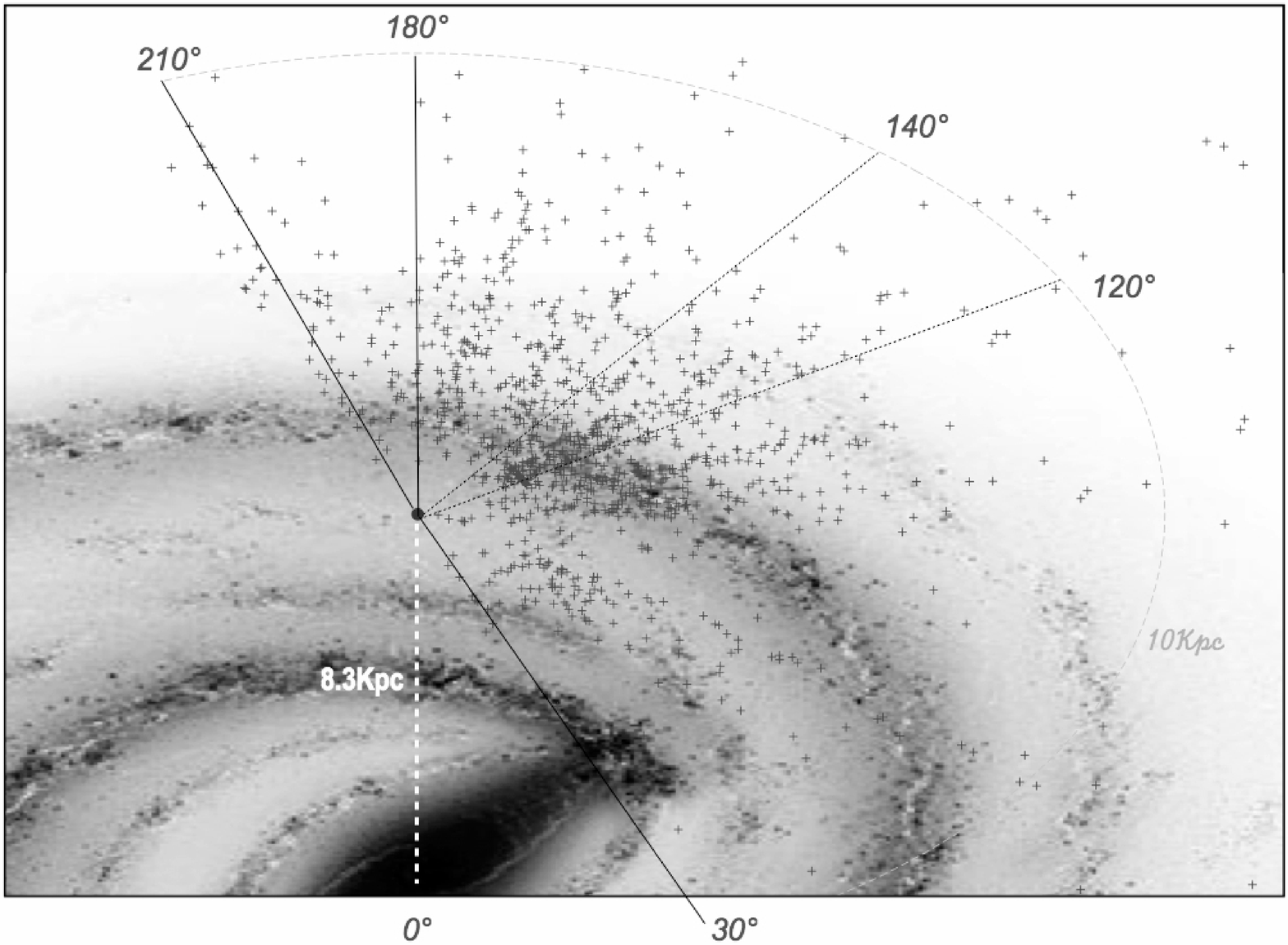}{ex_fig1}{Plot of the distances to CBe stars in the Northern Galactic plane region.The background image illustrates the results on the Galactic structure of the Spitzer space telescope.}

All the targets were first detected from IPHAS colour-colour diagrams as strong  H$\alpha$ emitters. However, the spectra of 83 stars (8\% of the sample) displayed the H$\alpha$ line in absorption. We interpret this fact as a phase transition from emission to absorption in the H$\alpha$ line. The change was produced during the interval between the photometric and spectroscopic data acquisition, which ranges from two to five years. This is compatible with the known timescales of this kind of phase transitions in CBe stars,  between four and fifteen years. For three of these stars we have two spectra at different epochs, in which the gradual transition from line emission to absorption can be traced, confirming the proposed scenario. The large amount of CBe stars showing the transition to a disk loss phase represents an opportunity to study the frequency of this phenomenon in CBe stars and its characteristic timescales. \par

\section{Discussion}
We present a catalogue containing spectral classification, astrophysical parameters and distances for a sample of  1135 CBe stars in the Northern Galactic plane. From Fig. 1 it is apparent  that a high percentage of the population lies on and around the region where the Perseus Arm is expected to be,  at distances between 2 and 4 Kpc.  The rest of the sample is scattered beyond this distance, without presenting a clear view on  the Outer Arm structure. Hence the present work can not firmly support the two arms structure of the Galaxy in the anticenter direction.\par

The density of the stellar population beyond the Perseus Arm seems to follow the same pattern of as almost exponential drop at all longitudes along the Northern Galactic plane. In the case that a large fraction of this population actually belongs to the Outer Arm, it would extend at least up to $210^o$. At some specific longitudes the presence of gaps in the stellar distribution is apparent. They are related to regions of high interstellar extinction, which are also seen in the 3D extinction map of the Northern Galactic plane elaborated from IPHAS data by \cite{Sale2014}.\par

In Fig. 2 we present an histogram of the CBe stellar population versus distance, in the $120^o-140^o$ galactic longitude range. The galactic structure in this region as been studied by \cite{Raddi2013}, who obtain results compatible with those presented in this work.\par

\articlefigure{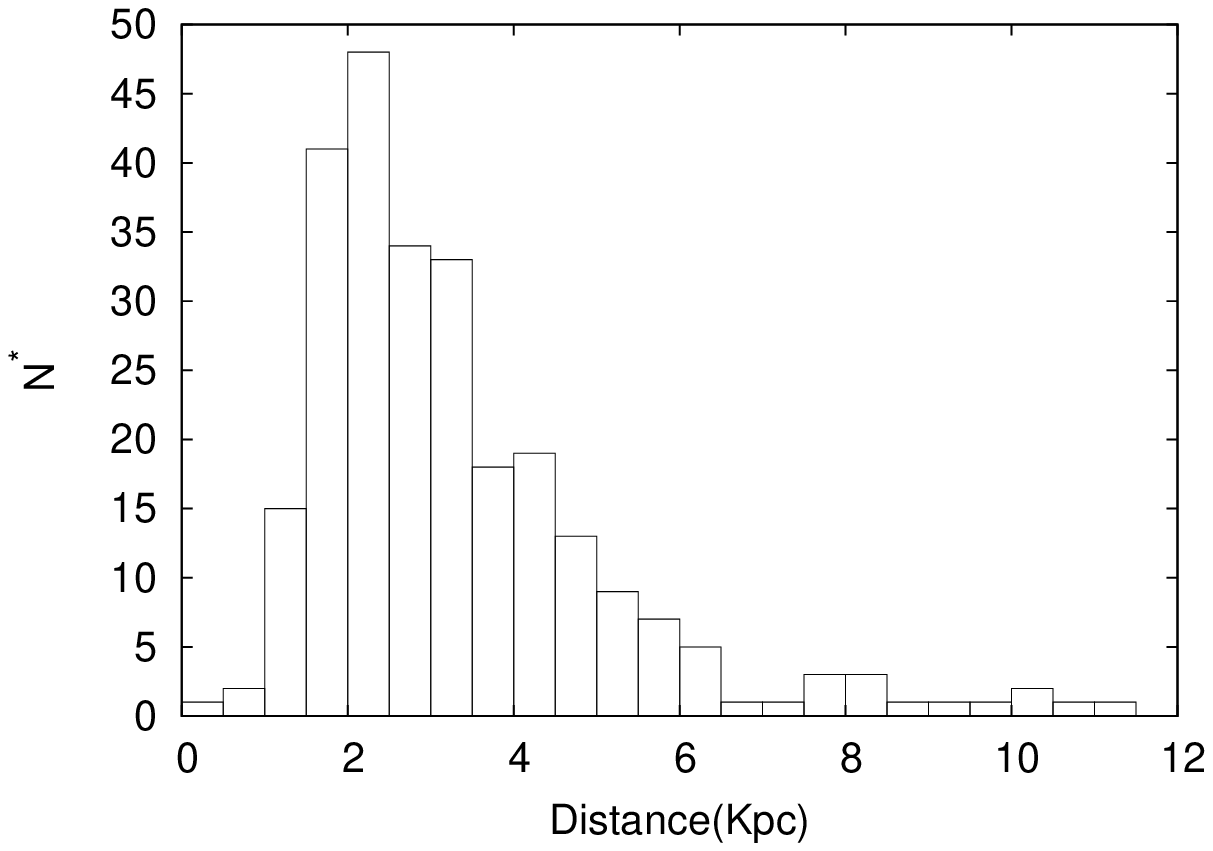}{ex_fig2}{Histogram of the CBe population versus the distance in the $120^o-140^o$ galactic longitude range.}

Finally, we note that a significant number of stars are detected at distances larger than 9 Kpc, and up to 11 Kpc. Current models predict that the  extension of the galactic disk in the anticenter direction is around, 8 Kpc., where the Outer Arm is supposed to end. We are investigating the farther targets one by one in order to examine at which level our results for each one is trustful, with the aim to study the outer limits of the galactic disk.

\section{Question}

 \textbf{Okazaki A. :} You mentioned that 8$\%$ of population changed from emission to
   absorption during 2-4 years. This seems to suggest that the 
   time-scale of changing from the Be phase to the normal B phase is
   30-40 yr. Are there similar number of stars that changed from
   absorption to emission during this period?\\
\textbf{Gkouvelis L. :} Unfortunately not. IPHAS survey is detecting new H$\alpha$ emitters, from which at this work we study the CBe stars. Only the difference in observational date, between the photometry and spectroscopy, tell us that the phase is changed (except few ocasions that we have more than one spectrum). First we detect a bright   H$\alpha$  emitter photometrically and after we classify those objects as B type stars in non-emission phase spectroscopically.





\clearpage 

\acknowledgements 
This work makes use of data obtained as part of IPHAS carried out at the Isaac Newton Telescope (INT). The INT is operated on the island of La Palma by the Isaac Newton Group in the Observatorio del Roque de los Muchachos of the Instituto de Astrofisica de Canarias. All IPHAS data are processed by the Cambridge Astronomical Survey Unit, at the Institute of Astronomy in Cambridge. The low-resolution spectra were obtained at the FLWO-1.5m with FAST, which is operated by Harvard-Smithsonian Center for Astrophysics.
In particular we want to thank Perry Berlind and Mike Calkins for their role in obtaining most of the FLWO-1.5m/FAST data. 
The work of L.G. and J.F. is supported by the Spanish Ministerio de Econom�a y Competitividad, and FEDER, under contract AYA2010-18352.





\bibliography{author}



%

\end{document}